\documentclass[a4paper]{jpconf}
\usepackage{graphicx}
\usepackage{physics}
\usepackage{color}
\usepackage{xcolor}
\usepackage{hyperref}
\usepackage{amssymb}
\begin{document}
\title{The quantum harmonic oscillator in a dissipative bath of anyon pairs}

\author{N-H Meyer\textsuperscript{1,a,*}, A Pelster\textsuperscript{2,b}, and M Thorwart\textsuperscript{1,c}}
\address{\textsuperscript{1}I.\ Institut für Theoretische Physik, Universität Hamburg, Notkestr.\ 9, 22607 Hamburg, Germany}
\address{\textsuperscript{2} Fachbereich Physik und Forschungszentrum OPTIMAS, Rheinland-Pfälzische Technische Universität Kaiserslautern-Landau, Erwin-Schrödinger-Straße 46, 67663 Kaiserslautern, Germany}
\address{\textsuperscript{*} Author to whom any correspondence should be addressed.}
\ead{\textsuperscript{a}nils-henrik.meyer@uni-hamburg.de, \textsuperscript{b}axel.pelster@rptu.de, \textsuperscript{c}michael.thorwart@uni-hamburg.de}

\begin{abstract}
We generalize the formalism of open quantum systems to introduce anyon baths. In particular, we 
work out a dissipative anyon bath composed of independent pairs of one-dimensional Grundberg-Hansson harmonically bound anyons, which are characterized by one statistical parameter. Using a mapping of these anyons to a bosonic bath with rescaled oscillator frequencies, we show that the original bilinear system-bath coupling assumes a particular non-polynomial form. To determine the relaxation properties, we use the imaginary-time path integral formalism together 
with a  generalization of Wick's theorem in the form of a smearing formula. The latter allows to approximately calculate the anyon bath spectral density, which acquires a nontrivial temperature dependence. The corresponding relaxation dynamics of the dissipative harmonic oscillator in an anyon bath is found. Well defined limits are revealed for both low and high temperatures. Anyonic features turn out to be most pronounced in the regime of intermediate temperatures. \\[-1cm]
\end{abstract}

\section{Introduction}
When a quantum system is exposed to quantum statistical fluctuations generated by the inevitable physical environment, it experiences relaxation and and decoherence \cite{weiss2012quantum}. The nature of the environmental fluctuations, which is often a fluctuating force, and their statistical properties crucially determine how the system approaches its equilibrium or steady state at asymptotic times. Classical systems exposed to classical fluctuating forces are investigated since the discovery of Brownian motion. For open quantum systems, in particular for quantum many-body systems, the relaxation behavior depends on whether the quantum statistical fluctuations are bosonic or fermionic in nature.  The approach to equilibrium is commonly described theoretically by taking the source of environmental fluctuations explicitly into account in the form of typical system-bath models. Models of the environment build on free boson or fermion baths. Commonly, the interaction between the system and the fluctuations is reduced to a form being both linear in the system and in the bath degrees of freedom. The fluctuating forces originating from the environment are commonly characterized by the bath spectral density $J(\omega)$. The central system of interest assumes a well-defined steady state if the bath spectral density behaves as $J(\omega\to 0)\to 0$, i.e., vanishes smoothly in the low-frequency limit. 

In recent years, the rise of quantum simulators based on ultracold atom gases has revived the interest in anyons \cite{Quantencomputing}. While systems of many bosons follow the Bose-Einstein statistics and fermions adhere to the Fermi-Dirac statistics, anyons correspond to any statistics, thereby connecting bosons and fermions which both represent limiting cases of anyons. The concept of anyons has been initially proposed by Leinaas and Myrheim in the 1970s \cite{anyonleinaas} and only a few years later, Wilczek coined the term \cite{anyonname}. Anyons were first observed in two-dimensional (2D) space as excitations within the fractional quantum Hall effect \cite{FHE}. Recent results have shown the experimental realization of anyons in one dimension (1D), i.e., as low-energy excitations in the fermionic Hubbard model \cite{Hubbertanyon}. They even can be created by spin-charge separation in strongly interacting quantum gases \cite{bosonicanyons}. Hence, anyons have evolved from being purely theoretical physics constructs to real physical entities. 

Since the processes of relaxation and quantum decoherence of a quantum system are crucially determined by the quantum statistical properties of the environmental fluctuations and the well-understood bosonic or fermionic fluctuations are in this sense limiting cases, it is natural to ask the question how a quantum system approaches equilibrium when exposed to anyonic quantum statistical fluctuations. This is the starting point of the present work, aiming to fill this conceptual gap. A useful starting point is to investigate - for simplicity - a 1D model, with the quantum harmonic oscillator as the central system. Modeling 1D anyons and constructing an environment out of 1D anyons is not unique, since it is presently an open question whether different known variants of 1D anyons can be unified to a common root.

A principal property of quantum statistics is the indistinguishability of their constituents. Being classically counterintuitive, this  property has far reaching consequences, e.g., for the exchange statistics of fundamental particles. In spatial dimensions larger or equal than three, the statistical properties can be symmetric or anti-symmetric with respect to particle exchange, which corresponds to bosonic or fermionic statistics \cite{anyonleinaas}. In lower spatial dimensions, the $N-$particle configuration space may become topologically non-trivial and anyonic quasiparticles with exotic exchange statistics \cite{Pauli1940,Harshman2020,Harshman2021,Harshman2023} may arise. It was conjectured that such exotic quasiparticles would emerge exclusively in 2D   \cite{anyonleinaas} and their role in connection with the origin of the fractional quantum Hall effect, topological phase transitions, braiding schemes of non-abelian anyons and their implications for quantum computing was discussed \cite{Iqbal}.

In 1D, though, the generic configuration space has rather a trivial topology \cite{Harshman2020,Harshman2021,Harshman2023} and exchange statistics and interactions are inextricably intertwined  \cite{Batista2004}, because particles can either move collectively or can scatter from each other while trying to exchange. It was first realized by Leinaas and Myrheim in 1977 that strictly locally interacting particles are dual to free particles evolving in an ordered space and the interactions correspond to boundary conditions for the wave function at the interface of two consecutive spatial regions \cite{anyonleinaas}. Almost four decades later, an exact quantum many-body formalism for 1D anyons was derived  \cite{PosskePRB}. The formalism explores  the equivalence to the Lieb-Liniger model of locally interacting bosons, for which an interpretation in the anyonic context has been established, in formal analogy to the 2D anyons of Wilczek \cite{PosskePRB,Wilczek1982}.  For a system of two 1D Leinaas-Myrheim anyons in a harmonic potential, Grundberg and Hansson constructed a coherent state path integral \cite{GRUNDBERG_1995}. An su(1,1)-Holstein-Primakoﬀ transformation yields a coherent states path integral for an ordinary harmonic oscillator with a shifted energy. This shift coincides with the one obtained for anyons by other methods. In the present work, we use this type of 1D anyon pairs to construct a system-anyon-bath model, see below. 

A further definition of 1D anyons was given by Kundu \cite{Kundu1999ExactSolutionOfDoubleDeltaFunctionBoseGasThroughAnInteractingAnyonGas}, generalizing the Lieb-Liniger model by spurious inter-particle interactions, a current-density coupling, and particularly balanced coupling constants. In analogy to Leinaas and Myrheim, the theory was  reformulated in terms of free anyonic fields \cite{Kundu1999ExactSolutionOfDoubleDeltaFunctionBoseGasThroughAnInteractingAnyonGas}. In 2011, 1D anyons were defined on a discrete lattice instead, with deformed commutation relations 
\cite{KeilmannLanzmichMcCullockRoncaglia2011StatisticallyInducedPhaseTransitionsAndAnyonsIn1DOpticalLattices}.  
A dual description exists in terms of either bosonic or fermionic {\em parent particles\/}  with the usual commutation relations, where the statistical phase factor appears as a density-dependent Peierls phase in their kinetic energy in the Hamiltonian. 
Then, a Hubbard anyon at a site $j$ corresponds to a boson or a fermion, which then performs correlated hopping \cite{KeilmannLanzmichMcCullockRoncaglia2011StatisticallyInducedPhaseTransitionsAndAnyonsIn1DOpticalLattices,Greschner2014}.
The lattice formulation of anyons has triggered novel experiments to engineer anyonic models via Floquet driving \cite{KeilmannLanzmichMcCullockRoncaglia2011StatisticallyInducedPhaseTransitionsAndAnyonsIn1DOpticalLattices,Greschner2014,Greschner2015,Eckardt2016,Goerg2019,Greiner2023,Tarurell2022}, optical wave guides \cite{Longhi2012}, or Raman-assisted coupling \cite{Tarurell2022}.  
The dilute lattice anyon model is equivalent to the Kundu model to first order in the statistical parameter, and to a solvable delta-derivative anyon gas to second order \cite{BonkhoffPRL, Bonkhoff2022thesis, Batchelor2008b}. It generalizes the Kundu model to higher-order current-density interactions and periodic coupling constants that inherit the topological character of the phases. 

This connection relates the continuum limit of the lattice anyon-Hubbard model to the Kundu anyons and thus to Leinaas' and Myrheim's anyons. This categorization was extended to multi-flavored anyons as well \cite{SantosParaanKorepin2012QuantumPhaseTransitionInAMulticomponentAnyonicLiebLinigerModel}. In such anyonic mixtures, non-standard interactions arise between the constituents that affect the particle statistics non-perturbatively, thus providing access to fundamentally novel features. 

In this work, we choose Grundberg-Hansson anyon pairs \cite{GRUNDBERG_1995} and construct an environment of an infinite set of mutually noninteracting anyon pairs. As will become clear below, such a constructed anyon bath can be translated into the common formalism of harmonic system-bath models, apart from one non-trivial aspect. Usually, one would focus on a bilinear coupling between system and bath degrees of freedom while studying the dynamical properties of a quantum system in a bosonic or a fermionic heat bath \cite{weiss2012quantum}. When we start from a bilinear coupling between the central quantum system and the set of anyon pairs and then perform the su(1,1)-Holstein-Primakoﬀ transformation to  construct the coherent states path integral for the anyonic bath and the system-bath coupling term, we obtain an effective system-anyon-bath coupling operator which is highly non-linear in the transformed anyonic operators. Such a model with a nonlinear coupling cannot be treated further exactly. To proceed, however, we can use a smearing formula for bipartite path integrals with non-polynomial interactions between the constituents \cite{Kleinert_1998,direlac} which builds on a variant  of a cumulant expansion of the effective propagator to lowest nontrivial order in the coupling strength. As shown in Ref.~\cite{direlac}, the smearing formula represents a generalization of Wick’s theorem of decomposing correlation functions involving functions of the canonical 
variables. By this, we obtain again an effective bilinear system-bath coupling term, however, with strongly renormalized coupling constants. In particular, this leads to temperature-dependent effective coupling constants, which enter in the formalism of quantum dissipative theory, eventually producing a system-bath spectral density which becomes depending on the bath temperature. Such cases are known to arise in different circumstances with non-linear system-bath couplings to bosonic baths \cite{weiss2012quantum}. The concept to apply the smearing formula to non-polynomial system-bath interactions is not restricted to anyon baths and constitutes a general approach. 

Hence, starting from a quantum harmonic oscillator, which is bilinearly coupled to a bath of anyon pairs, we are able to construct a mapping to a quantum harmonic oscillator which is non-polynomially coupled to a bath of free bosons which have shifted energies. The energy shift depends on the statistical parameter of the anyons. Aiming to determine the bath spectral density of this anyon-pair bath, we employ the imaginary-time path integral formalism. A direct integration over the bath degrees of freedom is not straightforward due to the non-polynomial system-bath coupling. However, using the smearing formula as a result from a systematic cumulant expansion, a bilinear system-bath coupling operator can be constructed, however, with a nontrivial renormalization of the initially chosen bath spectral density. By this, an initially chosen Ohmic anyon bath spectral density is renormalized into a non-standard effective bath spectral density which even becomes temperature dependent. Nevertheless, the effective bath spectral density can be employed further in the real-time path integral formalism. As a paradigm example, we study the relaxation of a dissipative quantum harmonic oscillator in real time induced by the anyon-pair bath. We find that the relaxation rate is directly proportional to the anyon statistical parameter. 

The paper is structured as follows: We start by introducing the underlying model in Sec.\ \ref{sec;model}, where a more detailed explanation of the Grundberg and Hansson anyons is given. Through the construction of the Euclidean action for the system, the bath of anyon pairs, and the bilinear interaction term, we encounter a non-bilinear coupling between system and bath after the su(1,1)-Holstein-Primakoﬀ transformation, which turns the bath of anyon pairs into a bath of renormalized free bosons. We then analyze
the fluctuation characteristics of the system-anyon-bath model in Sec.\ \ref{fluc-char}. To this end we
determine at first the underlying influence functional within a cumulant expansion in Sec.\ \ref{sec;Hauptteil}.
To proceed with the so generated non-polynomial system-bath coupling, we make use of the smearing formula in Sec.\ \ref{smearing-formula}.
The resulting nontrivial effective spectral density is further analyzed in Sec.\ \ref{sec;Spektral}. In Sec.\  \ref{sec;Relaxation}, we use this effective spectral density in the real-time formalism to derive the 
relaxation dynamics of the quantum harmonic oscillator coupled to the bath of anyon pairs. Finally, the corresponding conclusions are drawn in Sec.\ \ref{sec;summary}. 
\section{The model}
\label{sec;model}
We consider a bosonic harmonic oscillator with the frequency of $\omega_0$ as the central system in a system-bath model and use the imaginary-time path integral formalism to determine the fluctuation characteristics for a binlinear coupling to a bath of anyon pairs. The corresponding Euclidean action of the oscillator is given by 
\begin{equation}
\mathcal{S}_{\rm S}[a(\tau),a^*(\tau)]= \int_0^{\hbar\beta}d\tau \biggl\{ \frac{\hbar}{2}\biggl[\Dot{a}^*(\tau)a(\tau)-a^*(\tau)\Dot{a}(\tau)\biggr]+\hbar\omega_0 a^*(\tau)a(\tau)\biggr\}\,.
\label{system}
\end{equation}
Here $a(\tau)$ ($a^*(\tau)$) denotes the trajectory in imaginary time $\tau$ corresponding to the bosonic annihilation (creation) operator $a$ ($a^\dagger$) of the harmonic oscillator. And $\beta=1/(k_{\rm B} T)$ stands for the inverse thermal energy with $T$ being the temperature and $k_B$ being the Boltzmann constant. 

\begin{figure}[]
	\centering
	\includegraphics[width=0.5\textwidth]{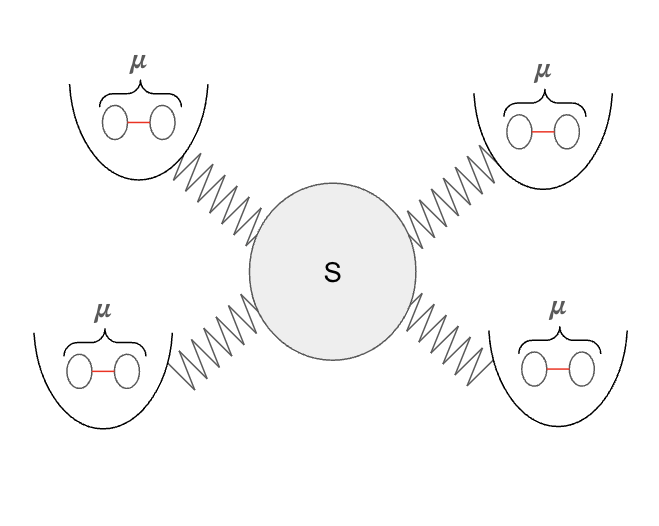}
	\caption{Scheme of the system-anyon-bath model constructed in this work.}
	\label{fig:modell}
\end{figure}

To construct the anyon bath, we consider one-dimensional anyons as introduced by Grundberg and Hansson \cite{GRUNDBERG_1995}. By trapping two anyons in a single harmonic potential, they can only be interchanged with each other and therefore get rid of the complicated exchange behavior of the anyons. Consequently, the Grundberg-Hansson anyons are characterized by only one statistical parameter $\mu$ and can be linked to the well-known statistical angle in the two-dimensional case \cite{GRUNDBERG_1995}. Here $\mu =1/4$ recovers the bosonic and $\mu =3/4$ the fermion case, respectively. To construct the bath of anyon pairs, we first start with a single pair of anyons, which is harmonically bound and construct their Euclidean path integral. To this end the coherent state path integral, summarized in real time in \ref{appendixa}, has to be Wick rotated. Furthermore, it is shown there how to use the su(1,1)-Holstein-Primakoﬀ transformation in order to map the path integral to that of a single bosonic harmonic oscillator with a rescaled oscillation frequency. For a collection of such anyon pairs, we add up many copies of single harmonically bound anyon pairs and obtain the Euclidean action of the  resulting bath of harmonically bound anyon pairs in the form  
\begin{equation}
\mathcal{S}_{\rm B}[b_{\omega}(\tau),b_{\omega}^*(\tau)]=2\mu \sum_{\omega}\int_0^{\hbar\beta}d\tau \biggl\{
\frac{\hbar}{2} \biggl[\Dot{b}_{\omega}^*(\tau)b_{\omega}(\tau)-b_{\omega}^*(\tau)\Dot{b}_{\omega}(\tau) \biggr]+ \hbar\omega b_{\omega}^*(\tau)b_{\omega}(\tau)\biggr\} \, .
\label{bath}
\end{equation}
Here, each harmonically bound anyon pair is characterized by its frequency $\omega$. The imaginary time paths $b_{\omega}(\tau)$ and $b_{\omega}^*(\tau)$ correspond to the bosonic annihiliation and creation operators $b_{\omega}$ and $b_{\omega}^\dagger$, respectively, which result from the corresponding anyonic operators after applying the su(1,1)-Holstein-Primakoﬀ transformation of the harmonically bound pair with frequency $\omega$. 
Since restrictions to the transformation apply \cite{GRUNDBERG_1995}, this mapping is valid only for the anyon statistical parameters $\mu > 1/2$. We therefore lose the ability to recover the purely bosonic case, but still include the fermionic case. We consider the situation that each of the bath anyon pairs couples individually with the  coupling strength $c_{\omega}$ to the central system oscillator, as illustrated in Fig.\ \ref{fig:modell}. The coupling constant $c_{\omega}$ in the coherent state representation is linked to the coordinate displacement coupling constant $c'_{\omega}$ via
\begin{equation}
    c_{\omega} = \frac{\hbar}{2\sqrt{m_{\omega}\omega M\omega_0}}\,c'_{\omega} \, ,
\end{equation}
where $M$ denotes the mass of the central system oscillator and $m_{\omega}$ stands for the mass of the anyon pair with  frequency $\omega$.
Consequently, the Euclidean action 
of the system-bath interaction follows
in the rotating wave approximation as   
\begin{equation}\label{euclidsysbath}
\mathcal{S}_{\rm SB}[b_{\omega}(\tau),b^*_{\omega}(\tau);a(\tau),a^*(\tau)]=2\mu\sum_{\omega} c_{\omega}\int_0^{\hbar\beta}d\tau \sqrt{1+|b_{\omega}(\tau)|^2}\biggl[a^*(\tau)b_{\omega}(\tau)+a(\tau)b^*_{\omega}(\tau)\biggr]\, .
\end{equation}
As apparent from Eq.\ (\ref{euclidsysbath}), the system-anyon-bath coupling is no longer bilinear in the respective degrees of freedom, but becomes non-polynomial due to the square root  in the anyon degrees of freedom. It is this feature, which prevents us from a one-to-one transfer of the results known for a conventional system-bosonic bath model to the case of anyons. As shown below, we develop a non-trivial approach on the basis of a smearing formula to tackle this non-polynomial system-bath interaction at least approximately. 
\section{Fluctuation characteristics of the system-anyon-bath model}\label{fluc-char}
An essential element for characterizing the thermodynamic properties of a quantum system is the underlying partition function. When dealing with a system-bath model, attention is focused on the reduced density matrix, which can be derived from the total density matrix of the system-bath model by eliminating the bath's degrees of freedom. Within the imaginary-time path-integral formalism, this is achieved by choosing the periodic imaginary paths of the bath part with equal start and end point and integrating over these points. To do so, we represent the 
partition function of the considered bosonic harmonic oscillator with the Euclidean action (\ref{system})
as a periodic coherent state path integral in imaginary time in the form  \cite{weiss2012quantum}
\begin{equation}
   Z = \oint \mathcal{D}\Bigl(a^*(\tau),a(\tau)\Bigr)
   e^{- \mathcal{S}_{\rm S}[a^*(\tau),a(\tau)]/\hbar}
   \,\mathcal{F}[a^*(\tau),a(\tau)]
   \, ,
    \label{equ;reddichte}
\end{equation}
where the influence functional is given by
\cite{weiss2012quantum}
\begin{equation}
    \mathcal{F}[a^*(\tau),a(\tau)]=
    \expval{e^{ -\mathcal{S}_{\rm SB}/\hbar}
    }_{\rm B}\, .
    \label{equ;influencefunc}
\end{equation}
Here, we use $\mathcal{S}_{{\rm SB}}$
as a short-cut notation for the Euclidean action in Eq.\ (\ref{euclidsysbath}),  describing 
the system-anyon bath coupling. Moreover, $\expval{\bullet}_{\rm B}$ stands for the harmonic expectation value evaluated over the bath degrees of freedom:
\begin{equation}
\expval{\bullet}_{\rm B}=\frac{1}{Z_{\rm B}}
\oint \mathcal{D}\Bigl(b^*_{\omega}(\tau),b_{\omega}(\tau)\Bigr) 
\bullet  \,  e^{-\mathcal{S}_{\rm B}[b^*_{\omega}(\tau),b_{\omega}(\tau)]/\hbar}\, ,
\end{equation}
with the partition function of the bath
\begin{equation}
Z_{\rm B}
=\oint \mathcal{D}\Bigl(b^*_{\omega}(\tau),b_{\omega}(\tau)\Bigr) 
 e^{-\mathcal{S}_{\rm B}[b^*_{\omega}(\tau),b_{\omega}(\tau)]/\hbar}\, .
\end{equation}
Thus, the influence functional, Eq.\ (\ref{equ;influencefunc}), 
contains all the information of the bath via Eq.\ 
(\ref{bath}) and its interaction with the system
according to Eq.\ (\ref{euclidsysbath}). 
\subsection{Influence functional}
\label{sec;Hauptteil}
To evaluate the influence functional, we expand the imaginary-time path integral (\ref{equ;influencefunc}) with respect to the interaction term and consider only the first non-trivial term. Due to the construction of the system-bath model, we find that only an equal number of trajectories $b_{\omega}(\tau)$ and $b_{\omega}^*(\tau)$ produce a non-zero contribution to the expansion, which leaves us in lowest order with
\begin{equation}
    \mathcal{F}[a^*(\tau),a(\tau)]=1+\frac{1}{2\hbar^2}
    \expval{ \mathcal{S}_{{\rm SB}} \mathcal{S}_{{\rm SB}}}_{\rm B} + \ldots \, .
    \label{perturb}
\end{equation}
As the bath, Eq.\ (\ref{bath}), consists of uncorrelated harmonically bound anyon pairs,
the first non-trivial influence of the bath is described by the  expectation value
\begin{equation}
\label{sbb}
    \expval{\mathcal{S}_{{\rm SB}} \mathcal{S}_{{\rm SB}}}_{\rm B}= (2\mu)^2\sum_{\omega}c_{\omega}^2\int_0^{\hbar \beta}d\tau_1 \int_0^{\hbar \beta}d\tau_2 
  \biggl[  a(\tau_2)a^*(\tau_1)P_{\omega}(\tau_1,\tau_2) +\left(1\leftrightarrow 2\right) \biggr]\, ,
\end{equation}
where we introduced the spectral correlation function
\begin{equation}\label{peval}
    P_{\omega}(\tau_1,\tau_2)=\expval{\sqrt{1+|b_{\omega}(\tau_1)|^2}\sqrt{1+|b_{\omega}(\tau_2)|^2} \, b_{\omega}(\tau_1)b^*_{\omega}(\tau_2)}_{\rm B}\,.
\end{equation}
Reexpressing the right-hand side of Eq.~(\ref{perturb}) in form of an exponential
\begin{equation}
    \mathcal{F}[a^*(\tau),a(\tau)]=
e^{-  \mathcal{S}_{{\rm infl}}  [a^*(\tau),a(\tau)]/\hbar}   \, ,
\label{cumulant}
\end{equation}
we obtain the lowest-order approximation
for the Euclidean influence action:
\begin{equation}
\mathcal{S}_{{\rm infl}}  [a^*(\tau),a(\tau)] = - \frac{1}{2\hbar}
\expval{ \mathcal{S}_{{\rm SB}} \mathcal{S}_{{\rm SB}}}_{\rm B} + \ldots
\,.
\label{influence-action}
\end{equation}
At this point, it remains in Eq.~(\ref{sbb}) to evaluate the expectation value defined in Eq.\ (\ref{peval}). This is not possible via
a straightforward application of Wick's theorem as the expectation value contains  the bath degrees of  freedom in a non-harmonic form.
This is the consequence of the nonlinear system-bath coupling, Eq.\ (\ref{euclidsysbath}), reflecting  the anyonic nature of the bath. 
\subsection{Smearing Formula}
\label{smearing-formula}
In case of Eq.~(\ref{peval}), where a harmonic expectation value for a non-polynomial function is to be calculated,
 a corresponding generalization of Wick's theorem is needed. This is provided by the so-called smearing formula \cite{Feynman,KleinertBook,Kleinert_1998}. 
The general idea is that the path expectation value boils down 
to convolute the non-polynomial functions with an appropriate Gau{\ss} function. This procedure yields for the spectral correlation function (\ref{peval})
\begin{eqnarray}
    P_{\omega}(\tau_1,\tau_2) &=&\int\frac{db_1 db^*_1 db_2db^*_2}{\pi^2 D} \,\sqrt{1+|b_1|^2}\,\sqrt{1+|b_2|^2}\,b^*_{1}b_{2} \nonumber\\
 \label{equ;kern}   
   && \times \exp{-\sum_{n,m =1}^{2} \,\frac{(-1)^{n+m}}{D} \, G_{\omega}(\tau_n,\tau_m)
    \, b_{n}b_{m}^*}  \, .
\end{eqnarray}
Here we have introduced the correlation functions
$G_{\omega}(\tau_n-\tau_m)=  \expval{b_{\omega}^*(\tau_n)b_{\omega}(\tau_m)}_{\rm B}$, whose evaluation yields
\begin{equation}
    G_{\omega}(\tau_1-\tau_2)= \Big[ \Theta(\tau_2-\tau_1)\,n_{\omega}^{\rm B} +\Theta(\tau_1-\tau_2)\left(1+n_{\omega}^{\rm B}\right)\Big] e^{-\omega(\tau_1-\tau_2)} \, ,
    \label{equ;greenfkt}
\end{equation}
with $\Theta(x)$ denoting the Heaviside function  and the Bose-Einstein distribution 
being given by
\begin{equation}
n_{\omega}^{\rm B}=\frac{1}{e^{\hbar\beta\omega}-1}\,.
\label{equ;Boseeinstein}
\end{equation}
Furthermore, $D$ represents the determinant of the 2x2 matrix of correlation functions $G_{\omega}(\tau_n,\tau_m)$ \cite{Kleinert_1998},  yielding 
\begin{equation}
    D=G^2_{\omega}(0)-G_{\omega}(\tau_1-\tau_2)G_{\omega}(\tau_2-\tau_1)\, .
    \label{equ;det}
\end{equation}
Inserting Eq.\ (\ref{equ;greenfkt}) into Eq.\ (\ref{equ;det}) we find
\begin{equation}
 D=1+n^{\rm B}_{\omega} \, ,
\end{equation}
which is independent of the imaginary times
$\tau_1, \tau_2$. 

In order to solve the integral in Eq.~(\ref{equ;kern}), we introduce the polar coordinate representation
of the complex numbers $b_{n}=r_n e^{i\phi_n}$ for $n=1,2$. With this the integrals with respect to the absolute values read
\begin{equation}
        \hspace*{-3mm}P_{\omega}(\tau_1, \tau_2)=
 \int_0^{\infty}   dr_1\int_0^{\infty} dr_2\,\left\{\frac{(r_1r_2)^2}{\pi^2 (1+n_{\omega}^{\rm B})}\,\sqrt{1+r_1^2}\sqrt{1+r_2^2} \,e^{-(r_1^2+r_2^2)} L(r_1,r_2)+\bigl[1\leftrightarrow2\bigr] \right\}\, ,
\end{equation}
and the remaining angular integrals lead to
\begin{equation}
    L(r_1,r_2) = \int_0^{2\pi}d\phi_1\int_0^{2\pi}d\phi_2 \,\exp\Bigl\{-i(\phi_1-\phi_2)+\frac{G_{\omega}(\tau_1-\tau_2)}{1+n_{\omega}^{\rm B}}\,r_1r_2\, e^{-i(\phi_1-\phi_2)}+\bigl[1\leftrightarrow2\bigr]\Bigr\}\,.
\end{equation}
Performing the integral over the phases $\phi_1, \phi_2$ by taking into account Ref.\ \cite[3.937(1), 3.937(2)]{gradshteyn}, we find 
\begin{equation}
        L(r_1,r_2) = 4\pi^2\frac{G_{\omega}(\tau_1-\tau_2)}{\sqrt{n_{\omega}^{\rm B}(1+n_{\omega}^{\rm B})}}\operatorname{I_1}
\left(2r_1r_2\sqrt{\frac{n_{\omega}^{\rm B}}{(1+n_{\omega}^{\rm B})^3}}\,\right)\,,
\end{equation}
where $\operatorname{I}_1(x)$ stands for the modified Bessel function.
Using a Taylor expansion of the modified Bessel function \cite[8.447(2)]{gradshteyn}
and after some substitution of variables, we are able to perform the integral  on the radial parts $r_1$ and $r_2$ \cite[3.383(4)]{gradshteyn}. In this way we find that the expectation value in Eq.\ (\ref{sbb}) is of the generic form
\begin{equation}
    \expval{ \mathcal{S}_{\rm SB} \mathcal{S}_{\rm SB}}_{\rm B}=\int_0^{\hbar \beta}d\tau_1 \int_0^{\hbar \beta}d\tau_2 \Bigl\{a(\tau_1)a^*(\tau_2) K (\tau_1-\tau_2)+\bigl[1\leftrightarrow 2\big]\Bigr\} \, ,
    \label{equ;kernel}
\end{equation}
where the bath influence is effectively described by the kernel
\begin{equation}
 K(\tau)=(2\mu)^2  \sum_\omega c_{\omega}^2
 \, k(\hbar \beta\omega)\, G_\omega(\tau)\, .
    \label{equ;kernel-full}
\end{equation}
It contains
the
temperature- and frequency-dependent parameter
\begin{equation}
k(\hbar \beta\omega)=\sum_{n=0}^{\infty}(n+1)\biggl[\frac{n_{\omega}^{\rm B}}{(1+n_{\omega}^{\rm B})^{3}}\biggr]^n\operatorname{U}^2\left(n + 2, n + \frac{7}{2},1\right) \, ,
    \label{equ;Konstantek}
\end{equation}
where $\operatorname{U}(a,b,x)$ denotes the confluent hypergeometric function of second kind. Inserting therein the Bose-Einstein distribution (\ref{equ;Boseeinstein}), we obtain concretely
\begin{equation}
    k(\hbar\beta\omega)=  \sum_{n=0}^{\infty}(n+1)2^n e^{-2n\hbar\omega\beta}\biggl(\cosh{\hbar\omega\beta}-1\biggr)^n\operatorname{U}^2\left(n + 2, n + \frac{7}{2},1\right)\,.
    \label{equ;Konstantek2}
\end{equation}
In passing, we note that the series 
\begin{equation}
     k(x)= \sum_{n=0}^{\infty}a_n
     = \sum_{n=0}^{\infty}(n+1) \biggl(e^{-3x}+e^{-x}-2e^{-2x}\biggr)^n\operatorname{U}^2\left(n + 2, n + \frac{7}{2},1\right)
\end{equation}
is convergent, which follows from the limit 
\begin{eqnarray}
     \lim_{n\to \infty}\left|\frac{a_{n+1}}{a_n}\right| & = & \lim_{n\to \infty}\left| 
     \frac{(n+2) \biggl(e^{-3x}+e^{-x}-2e^{-2x}\biggr)^{n+1}\operatorname{U}^2\left(n + 3, n + \frac{9}{2},1\right)}{(n+1) \biggl(e^{-3x}+e^{-x}-2e^{-2x}\biggr)^{n}\operatorname{U}^2\left(n + 2, n + \frac{7}{2},1\right)}\right|\nonumber \\
     & = & | e^{-3x}+e^{-x}-2e^{-2x}| \nonumber \\
     & < & 1 \, .
\end{eqnarray}
\subsection{Spectral density}
\label{sec;Spektral}
We can now physically interpret our result for the expectation value in Eq.\ (\ref{equ;kernel}) with the correlation function $K(\tau)$ from Eq.\ (\ref{equ;kernel-full}). Although it describes a  harmonic oscillator coupled nonlinearly to the anyon bath, it formally resembles the corresponding expression for a quantum system, which is bilinearly coupled to a bosonic heat bath. 
The only difference is the temperature- and frequency-dependent prefactor given in Eq.\ (\ref{equ;Konstantek2}),
 which roots in the non-bilinear form of the system-bath coupling. Since this prefactor is
independent of the imaginary time, the equilibrium Green's function of the known harmonic oscillator
 $G_{\omega}(\tau)$ from Eq.\ 
 (\ref{equ;greenfkt})
contains the only imaginary-time dependence. Therefore, the fluctuation-dissipation theorem for the bosonic harmonic oscillator  \cite{weiss2012quantum}
\begin{equation}
G_{\omega} (- \tau) = G_{\omega} (\hbar \beta - \tau)\, , \hspace*{1cm} 0 < \tau < \hbar \beta \, ,
\end{equation}
implies a corresponding fluctuation-dissipation theorem 
\begin{equation}
K (- \tau) = \, K (\hbar \beta - \tau)\, , \hspace*{1cm} 0 < \tau < \hbar \beta \, , 
\end{equation}
for the  system-anyon bath model.

Based on this insight we can now
 read off the anyon bath spectral density. To this end we use the fact that the correlation function in Eq.\ (\ref{equ;kernel-full})
 is linked to the bath spectral density $J(\omega)$ via the relation \cite{weiss2012quantum}
\begin{equation}
    \begin{split}
        K(\tau) = \int_0^{\infty}d\omega' J(\omega')G_{\omega'}(\tau) \, ,
    \end{split}
\end{equation}
where $G_{\omega}(\tau)$ denotes the unperturbed thermal Green's function of a single harmonic oscillator. In the considered model 
the system-bath coupling constants and the anyonic bath oscillator frequencies are assumed to be given by the initial anyon bath spectral density \cite{weiss2012quantum}
\begin{equation}
   J_0(\omega')= (2\mu)^2\sum_\omega c_\omega^2 \delta(\omega-\omega') \, .
\end{equation}
Therefore, we find for the spectral density of the effective bosonic bath the general form 
\begin{equation}
    J(\omega, \beta) = k(\hbar\beta \omega)J_0(\omega) \, ,
    \label{equ;Spektraldichtefinal}
\end{equation}
with the  parameter  $k(\hbar\beta \omega)$  defined in Eq.\  (\ref{equ;Konstantek2}).
Because the relation (\ref{equ;Spektraldichtefinal}) allows us to perform the continuum limit of infinitely many bath anyons,
it represents the major insight of the present work. 

 Due to the non-bilinear coupling of the anyon bath to the system, the effective spectral density in Eq.\ (\ref{equ;Spektraldichtefinal}) turns out to depend on temperature. Let us therefore analyze
 the rescaling factor $k(\hbar\beta\omega)$ in the low- and high-temperature limit, respectively. A straightforward analysis yields in the low-temperature limit 
\begin{equation}
    k(\hbar\beta \omega) =\operatorname{U}^2\left(2, \frac{7}{2},1\right) + 2  \operatorname{U}^2\left(3, \frac{9}{2},1\right) e^{-\hbar \omega \beta}+ \ldots  \,  , \hspace*{1cm}  \hbar \beta \omega\to \infty \, ,
    \label{equ;inf}
\end{equation}
and, correspondingly, in the high-temperature limit
\begin{equation}
    k(\hbar \beta \omega) =\operatorname{U}^2 \left(2, \frac{7}{2},1\right) +  \operatorname{U}^2\left(3, \frac{9}{2},1\right) \left(\hbar \beta \omega\right)^2+ \ldots \, , \hspace*{1cm}
   \hbar \beta \omega\to 0\, .
    \label{equ;kwnull}
\end{equation}
Apparently, the anyonic nature of the bath survives in both limits.

\begin{figure}[t]
	\centering
\includegraphics[width=0.75\textwidth]{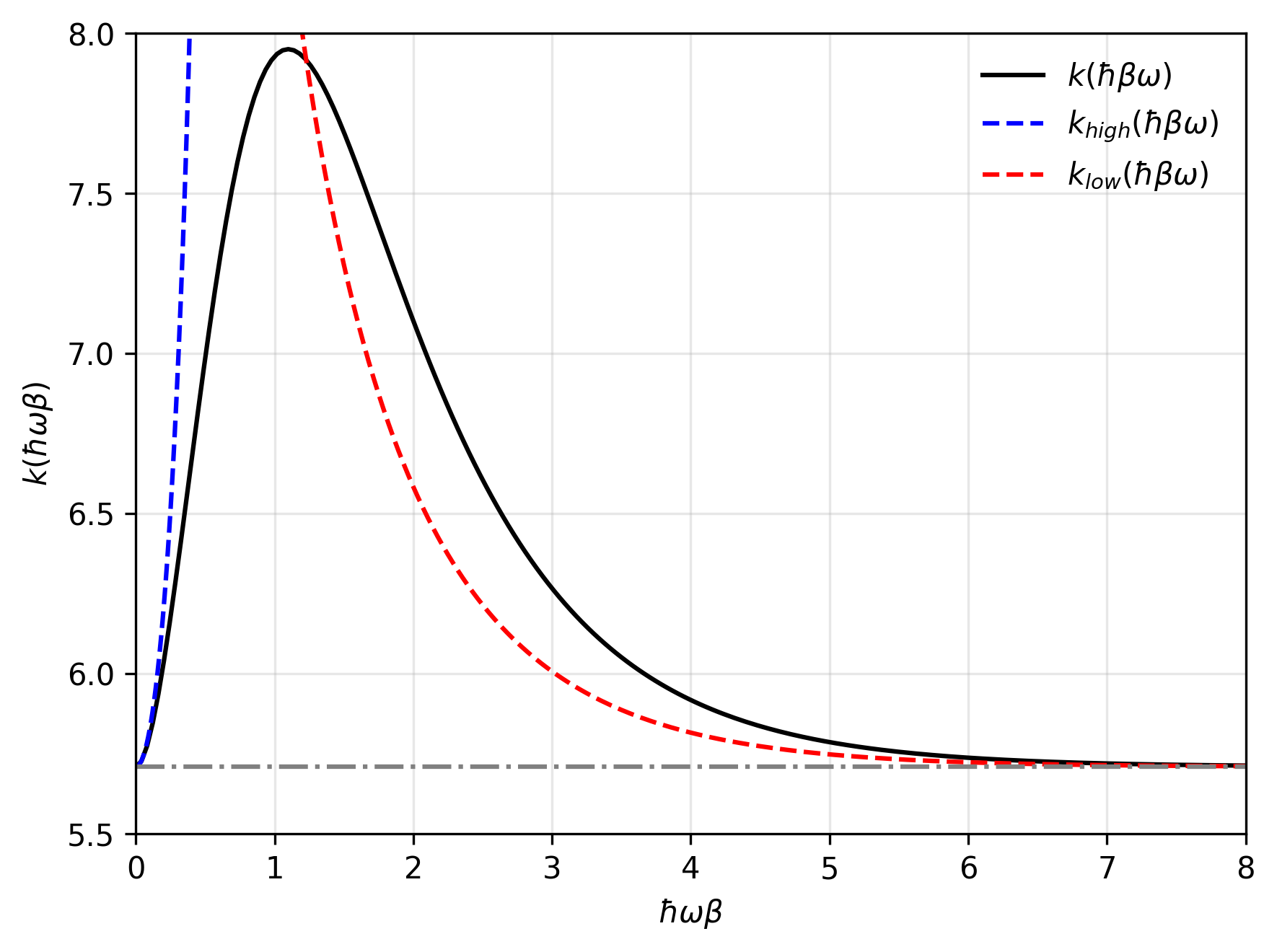}
	\caption{Temperature-dependent factor, Eq.\ (\ref{equ;Konstantek2},  of the effective spectral density in Eq.\ (\ref{equ;Spektraldichte}). The 
     red dashed line indicates the limit $\hbar \beta \omega\to \infty$ in Eq.\ \ref{equ;inf}, while the
    blue dashed line marks the limit of $\hbar \beta \omega\to 0$ in Eq.\ \ref{equ;kwnull}. The dash-dotted horizontal line stands for the common constant of both limits.}
	\label{fig:Spektraldichte}
\end{figure}

 In order to illustrate the impact of the anyonic character of the quantum statistical fluctuations, we choose an Ohmic bath spectral density $J_0(\omega)=(2\mu)^2 M\gamma \omega e^{-\frac{\omega}{\omega_c}}$ for the initial anyon bath. Here, $\gamma$ denotes the damping constant, $M$ is the mass of the system oscillator, and $\omega_c$ is a high-frequency cutoff. In this work, we consider the (standard) case when $\omega_c$ is the largest frequency in the model. With this the effective spectral density, Eq.\ (\ref{equ;Spektraldichtefinal}),  reduces to 
\begin{equation}
    J(\omega,\beta) = (2\mu)^2 M\gamma \omega\, k(\hbar \beta\omega) \, .
    \label{equ;Spektraldichte}
\end{equation}
The functional dependence of the parameter $k(\hbar \beta\omega)$ in Eq.\ (\ref{equ;Konstantek2})
on temperature  is
shown in units of $\hbar \omega_0/k_{\rm B}$ in Fig.~\ref{fig:Spektraldichte}.
It turns out to be peaked around $\hbar \beta \omega \approx 1$.
Both in the low- and the high-temperature limit, the spectral density in Eq.\ (\ref{equ;Spektraldichte}) turns out to converge to the same constant value $(2\mu)^2 M\gamma \omega\operatorname{U}^2(2,7/2,1)$, thus recovering structurally an ordinary Ohmic spectral density. 

\begin{figure}[t]
	\centering
	\includegraphics[width=0.75\textwidth]{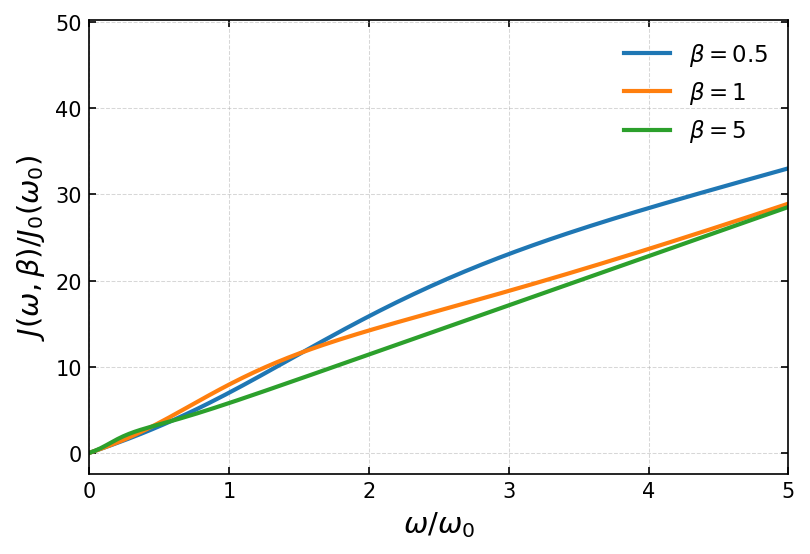}
	\caption{Comparison of the anyonic spectral density $J(\omega,\beta)$ for different temperatures, expressed in units of $\hbar\omega_0$, scaled with the anyon bath spectral density $J_0(\omega_0)$, as a function of $\omega /\omega_0$.}
	\label{fig:Spektraldichtetemperatur}
\end{figure}

In Fig.\ \ref{fig:Spektraldichtetemperatur}, we show the resulting total spectral density $J(\omega,\beta)$, Eq.\ (\ref{equ;Spektraldichte}), for different choices of the temperature. We observe the deviation from a pure Ohmic form in the region of the frequency $\omega_0$ of the system oscillator. Since only a pure Ohmic form leads to Markovian relaxation dynamics \cite{weiss2012quantum},
we conclude that the anyon bath induces a 
non-Markovian dynamics.

\section{Relaxation dynamics}
\label{sec;Relaxation}
The effective spectral density, Eq.\ (\ref{equ;Spektraldichte}), allows us to study numerous dynamical properties of the central quantum system. For instance, following Refs.\ \cite{weiss2012quantum,Ingold}, the relaxation dynamics of a quantum dissipative harmonic oscillator is determined by the real-time damping kernel $\gamma(t)$, which is directly related to the bath spectral density $J(\omega,\beta)$ according to 
\begin{equation}
  \gamma(t)=\Theta(t)\,\frac{2}{M\pi}\int_0^{\infty}d\omega \,\frac{J(\omega,\beta)}{\omega}\,\cos{\omega t}\, .
  \label{gammavont}
\end{equation}
To this end the Laplace transform 
\begin{equation}
    \hat{\gamma}(z) = \int_0^{\infty}dt \, e^{-zt}\,\gamma(t)
    \label{equ;gammaz}
\end{equation}
of the damping kernel is needed. Inserting the spectral density from Eq.\ (\ref{equ;Spektraldichte})
together with Eq.\ (\ref{equ;Konstantek2}) 
into Eq.\ (\ref{gammavont}) and performing the integral over real time $t$ in Eq.\ (\ref{equ;gammaz}), we arrive at
\begin{equation}
    \hat{\gamma}(z) =\frac{8\mu^2 \gamma}{\pi}\sum_{n=0}^{\infty}(n+1)2^n\operatorname{U}^2\left(n + 2, n + \frac{7}{2}, 1\right)\int_0^{\infty}d\omega \, e^{-2n\omega\hbar \beta}\biggl(\cosh{\omega\hbar\beta}-1\biggr)^{n}\frac{z}{z^2+\omega^2} \, .
    \label{equ;gamma(z)einge}
\end{equation}
Also this damping kernel has well-defined limits for low and high temperatures, respectively. In the low-temperature limit one can evaluate the appearing frequency integral with the help of \cite[3.542(1)]{gradshteyn}, yielding
\begin{equation}
        \hat{\gamma}^{\rm low}(z) =\gamma_{\mu}+\frac{T}{z}\,c+ \ldots \,.
    \label{equ;gammalow}
\end{equation}
Here the temperature-independent constant 
\begin{equation}
    \gamma_{\mu} = (2\mu)^2 \gamma\operatorname{U}^2\left(2,\frac{7}{2},1\right) 
    \label{constant}
\end{equation}
contains an anyonic characteristic due to the appearance of $\mu$. Moreover, the
linear temperature correction in Eq.\ (\ref{equ;gammalow}) is determined by the 
constant 
\begin{equation}
        c =\frac{8\mu^2 \gamma}{\pi}\sum_{n=1}^{\infty}(n+1)\operatorname{U}^2\left(n + 2, n + \frac{7}{2}, 1\right)\operatorname{B}(n,2n+1) =25.171\frac{\mu^2 \gamma}{\pi}\, ,
    \label{equ;gammalow2}
\end{equation}
where $\operatorname{B}(x,y)$ stands for the beta function \cite[8.384(1)]{gradshteyn}. Correspondingly, the high-temperature limit leads to a Gaussian integral, which is straightforwardly evaluated to 
\begin{equation}
        \hat{\gamma}^{\rm high}(z) =\gamma_{\mu}+\frac{z}{T}\,d \, .
    \label{equ;gammahigh}
\end{equation}
Here, the temperature correction depends on the constant
\begin{equation}
        d =\frac{8\mu^2 \gamma}{\pi}\sum_{n=1}^{\infty}(n+1)\operatorname{U}^2 \left(n + 2, n + \frac{7}{2}, 1\right)\,\frac{(2n-2)!}{(2n)^{2n-1}}=34.918\frac{\mu^2 \gamma}{\pi} \, .
    \label{equ;gammahigh2}
\end{equation}
Note that also
the damping kernel converges to the same constant in Eq.\ (\ref{constant}) in the limits of both low and high temperatures. Based on the Laplace transform of the damping kernel in Eq.\ (\ref{equ;gamma(z)einge})
the dissipative real-time dynamics of the oscillator displacement is determined by the roots of the  equation \cite{Ingold}
 \begin{equation}
 \label{equation}
        z^2+\omega_0^2+\hat{\gamma}(z)z=0 \, .
\end{equation}
In the limit of low temperatures, inserting Eq.\ 
(\ref{equ;gammalow}) 
into Eq.\ (\ref{equation}) yields the two roots
\begin{equation}
            \lambda^{\rm low}_{1,2} = -\frac{ \gamma_{\mu} }{2} 
            \pm i\sqrt{\omega_0^2+cT-\frac{\gamma_\mu^2}{4}} \, . \label{equ;Nllsttief}        
\end{equation}
Correspondingly, in the limit of high temperatures,  solving Eq.\ (\ref{equation}) by taking into account Eq.\  (\ref{equ;gammahigh}), we find the roots 
\begin{equation}
            \lambda^{\rm high}_{1,2} =\frac{1}{1+d / T} \left[- \frac{\gamma_{\mu}}{2} \pm i
        \sqrt{\omega_0^2\left(1+\frac{d}{T}\right)-\frac{\gamma_{\mu}^2}{4}}\,
            \right] \, .           \label{equ;Nllsthoch}
\end{equation}
As mentioned above, we consider the standard case of a large cut-off frequency $\omega_c \gg\omega_0$. Here, we exploit the fact \cite{weiss2012quantum} that  the expectation value $\expval{q}_t$ of the oscillator displacement is independent of $\omega_c$ and therefore justifies to take the limit $\omega_c \to \infty$.  In contrast, this does not hold for the expectation value $\expval{p}_t$ of the momentum, which depends sensitively on $\omega_c$, but is not considered here. Thus, in both limits, 
the relaxation dynamics of the displacement $\expval{q}_t$ is of the form
\begin{equation}
    \expval{q}_t=\expval{q}_0e^{-\Gamma t}\cos \Omega t \, ,
    \label{equ;Ortserwartung}
\end{equation}
where $\expval{q}_0$ denotes a given initial preparation  of the displaced system oscillator.
Moreover, $\Gamma$ and $\Omega$ stand for the corresponding damping rate and shifted frequency.
In the low-temperature limit, we get
\begin{equation}
\Gamma^{\rm low} = \frac{\gamma_{\mu}}{2}\, , \hspace*{1cm}
\Omega^{\rm low} = \sqrt{\omega_0^2+cT-\frac{\gamma_{\mu}^2}{4}}\, ,
    \label{equ;Ortserwartung}
\end{equation}
while the high-temperature limit yields
\begin{equation}
\Gamma^{\rm high}= \frac{\gamma_{\mu}}{2\left( 1+d / T \right)}       \, , \hspace{1cm}
     \Omega^{\rm high} = \sqrt{\frac{\omega_0^2}{1 + d/T}-\frac{\gamma_{\mu}^2}{4(1+d / T)^2}} \, .
\end{equation}
Noticeably, in both cases these parameters have a similar structure as for bosonic system
with Ohmic damping \cite{weiss2012quantum}. But the striking difference is that relaxation rate and oscillation frequency become generically temperature dependent, which can be traced back via Eqs.\ (\ref{equ;gammalow2}) and (\ref{equ;gammahigh2}) to the anyon parameter $\mu$.

\begin{figure}[t]
	\centering
	\includegraphics[width=0.65\textwidth]{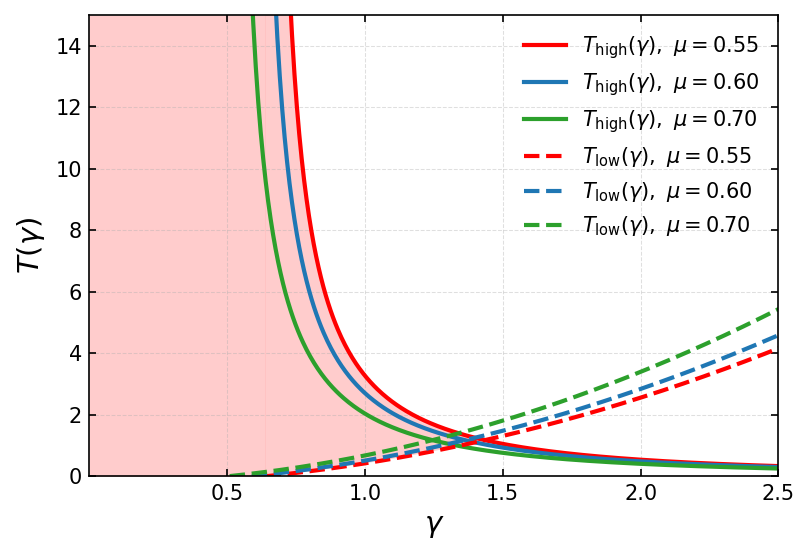}
	\caption{Phase diagram of the coherent-incoherent dynamical regimes in the temperature $T$–damping constant $\gamma$ plane for various anyonic parameters $\mu$ as indicated. 
    The separation lines are defined by the roots of the radicand of Eqs.\ (\ref{equ;Nllsttief}) [low temperature, dashed lines] and (\ref{equ;Nllsthoch}) [high temperature, solid lines]. The shaded region represents the coherent regime for $\mu = 0.55$ case (shown in red). The exact separatrix for a general value of $T$ cannot be evaluated analytically.}
	\label{fig:TvonGamma}
\end{figure}

Finally, we investigate the transition between coherent dynamics, with a decaying oscillatory behavior, and incoherent dynamics, with a purely exponential decay. The two regions are separated by the condition that the imaginary parts of the eigenvalues in Eqs.\ (\ref{equ;Nllsttief}) and (\ref{equ;Nllsthoch}) vanish. This can be evaluated analytically only in the mentioned limits. We find a resulting phase diagram for the coherent/incoherent dynamical regimes, shown in Fig.\ \ref{fig:TvonGamma} for different values of the statistical parameter $\mu$ as indicated. 

\section{Conclusions}
\label{sec;summary}
The spectral characteristics of environmental fluctuations crucially determine the nonequilibrium dynamics of open quantum systems. While bosonic and fermionic environments are well established, anyon baths have not been considered before. In this work, we introduce a specific system-anyon-bath model in which the bath is composed of Grundberg-Hansson harmonically bound anyon pairs. We use the appealing property of a Grundberg-Hansson one-dimensional anyon pair that it can be mapped to a single bosonic harmonic oscillator with a rescaled frequency. The rescaling depends on the anyon statistical parameter. For the conventional bilinear system-bath coupling to the displacement of the anyon pair, the mapping to effective boson modes leads to a particular effective system-bath coupling in the bosonic framework which is non-polynomional in the anyon degrees of freedom. 

To determine the anyon bath spectral density, we use the paradigmatic example of the harmonic oscillator as the system and employ the imaginary time path integral formalism. The non-polynomial system-anyon bath coupling prevents us from a direct integration over the bath degrees of freedom. Instead, we use a smearing formula as a generalization of Wick's theorem and evaluate approximately the fluctuation effects of the non-polynomial coupling. This leads to an unconventional environmental spectral density which, in particular, becomes temperature-dependent. Interestingly, the temperature dependence has well defined limits for low and high temperatures. 

Having derived the anyon bath spectral density, we use it to study the real-time relaxation dynamics of the dissipative harmonic oscillator for the case of an initial Ohmic anyon bath. We show that the anyon bath induces a temperature dependent enhancement of the fluctuations as compared to the bosonic case. The model allows us to derive analytic results for the relaxation rate and the oscillation frequency of the dissipative harmonic oscillator. We also obtain a parameter-dependent separatrix between the regimes of coherent and incoherent relaxation of the oscillator. 

The model of the bath consisting of Grundberg-Hansson harmonically bound anyon pairs is particular and could be modified in various directions. For instance, one could also construct an anyon bath starting from the one-dimensional anyon-Hubbard model \cite{BonkhoffPRL} and obtain a different model of the bath. Interesting questions arise, e.g., about the equivalence and the universality of such anyon bath models which are left to future investigations.  

\ack
We acknowledge funding by the Deutsche Forschungsgemeinschaft (DFG, German
Research Foundation) – Projektnummer 547748180.

\begin{appendix}
\section{Coherent state path integral of a Grundberg-Hansson anyon pair coupled to a bosonic ocillator}\label{appendixa}
\renewcommand{\theequation}{A.\arabic{equation}}
\setcounter{equation}{0}

Here, we demonstrate how to construct the coherent state path integral for the Grundberg-Hansson anyon pair as introduced in Ref.\ \cite{GRUNDBERG_1995} and relate it to that of a bosonic harmonic oscillator by means of an su(1,1)-Holstein-Primakoff transformation. Moreover, we explicitly show this transformation also for a Grundberg-Hansson anyon pair coupled to another harmonic oscillator. 
\subsection{Algebra su(1,1)}
We begin with a concise overview of the su(1,1) algebra underlying the SU(1,1) symmetry group and how to construct the associated coherent states which form the basis for the Grundberg-Hansson anyon pairs. For more technical details, we refer to Ref.\ \cite{Perelomov}. We start from the three generating operators of the su(1,1) algebra, i.e., $\operatorname{K}_1$, $\operatorname{K}_2$, and $\operatorname{A}$, which are defined by the commutation relations
\begin{equation}
\label{comm}
\Bigl[\operatorname{K}_1,\operatorname{K}_2\Bigr] = -i\operatorname{A}\, , \qquad \Bigl[\operatorname{K}_2,\operatorname{A}\Bigr] = i\operatorname{K}_1\, , \qquad \Bigl[\operatorname{A},\operatorname{K}_1\Bigr] = i\operatorname{K}_2\, .
\end{equation}
By introducing the creation and annihilation operators $\operatorname{K}_{+}$ and $\operatorname{K}_{-}$ as linear combinations of $\operatorname{K}_1$ and $\operatorname{K}_2$ in the form 
\begin{equation}
        \operatorname{K}_{\pm} = \pm i\Bigl(\operatorname{K}_1 \pm i\operatorname{K}_2\Bigr),
        \label{equ;su11leiter}
\end{equation}
the commutation relations (\ref{comm}) become
\begin{equation}
\Bigl[\operatorname{A},\operatorname{K}_{\pm}\Bigr] = \pm \operatorname{K}_{\pm}\, , \qquad \Bigl[\operatorname{K}_-,\operatorname{K}_+\Bigr] = 2\operatorname{A} \, .
    \label{equ;kommutatorsu11}
\end{equation}
They act on a number state $\ket{n,\mu}$, with $\mu$ being the statistical parameter, and $n \in \mathbb{N}_0$ according to  
\begin{eqnarray}
\label{generator1}
        \operatorname{A}\ket{n,\mu} &= & (n+\mu)\ket{n,\mu}\, ,\\
        \label{generator2}
        \operatorname{K}_+\ket{n,\mu} &= &\sqrt{(n+1)(n+2\mu)}\ket{n+1,\mu}\, ,\\
\label{generator3}        
        \operatorname{K}_-\ket{n,\mu} &=& \sqrt{n(n-1+2\mu)}\ket{n-1,\mu}\, .
\end{eqnarray}
Next, one defines the su(1,1) coherent states 
\begin{equation}
    \ket{\zeta} = C e^{\zeta\operatorname{K}_+} \ket{0,\mu},
\end{equation}
where $\zeta$ denotes a complex number and $C$ represents a normalization constant. By performing a Taylor expansion of the exponential function and repeatedly applying Eq.\ (\ref{generator2}), one obtains
\begin{equation}
    \ket{\zeta} = C \sum_{n=0}^{\infty} \frac{\zeta^n}{n!} \sqrt{\prod_{j=0}^n(j+1)(j+2\mu)} \ket{n,\mu} \, .
\end{equation}
Using the Pochhammer symbol \cite{gradshteyn}, this expression can be rewritten as
\begin{equation}
    \ket{\zeta} = C \sum_{n=0}^{\infty} \frac{\zeta^n}{n!} \sqrt{\frac{2\mu(2\mu+1)_n}{n!}} \ket{n,\mu}  
\end{equation}
or in terms of the Gamma function $\Gamma(x)$,
\begin{equation}
    \label{coh-st1}
    \ket{\zeta} = C \sum_{n=0}^{\infty} \zeta^n \sqrt{\frac{\Gamma(2\mu+n)}{n!\, \Gamma(2\mu)}} \ket{n,\mu} \, .
\end{equation}
Apparently,  the scalar product of two distinct coherent states, Eq.\ (\ref{coh-st1}), is non-zero and becomes
\begin{equation}
   \bra{\zeta}\ket{\zeta}= C^2\sum_{n=0}^{\infty} |\zeta|^{2n} \frac{(2\mu+1)_n}{n!}\, .
\end{equation}
Evaluating the series yields
\begin{equation}
    \bra{\zeta}\ket{\zeta}=\frac{C^2}{(1-|\zeta|^2)^{2\mu}}\, .
\label{equ;skalarproduktSU11}
\end{equation} 
To ensure normalization of the coherent states, the normalization constant must be chosen as
\begin{equation}
    C=(1-|\zeta|^2)^{\mu},
\end{equation}
which finally gives 
\begin{equation}
    \label{coh-st}
    \ket{\zeta} = (1-\zeta)^{\mu} \sum_{n=0}^{\infty} \zeta^n \sqrt{\frac{\Gamma(2\mu+n)}{n!\, \Gamma(2\mu)}} \ket{n,\mu} \, .
\end{equation}%
These coherent states of Eq.\ (\ref{coh-st})  form an overcomplete basis and have the
completeness relation
\begin{equation}
    1=\int_{|\zeta|<1} \frac{d\zeta d\zeta^*}{\pi}\frac{2\mu-1}{(1-|\zeta|^2)^2}\ket{\zeta}\bra{\zeta}\, ,
    \label{equ;su11iden}
\end{equation}
which restricts the applicability to statistical parameters $\mu > \frac{1}{2}$.
In the basis of the coherent states, Eq.\  (\ref{coh-st}), the basis generators of the su(1,1) algebra turn out to have the matrix elements
\begin{eqnarray}
&\bra{\zeta}\operatorname{A}\ket{\zeta'}
    =\mu\, {\displaystyle \frac{(1-|\zeta|^2)^{\mu}(1-|\zeta'|^2)^{\mu}}{(1-\zeta^*\zeta')^{2\mu}}\,\frac{1+\zeta^*\zeta'}{1-\zeta^*\zeta'}}\, ,\\
    &\bra{\zeta}\operatorname{K_-}\ket{\zeta'}
    =2\mu\,\displaystyle{\frac{(1-|\zeta|^2)^{\mu}(1-|\zeta'|^2)^{\mu}}{(1-\zeta^*\zeta')^{2\mu}}\,\frac{\zeta'}{1-\zeta^*\zeta'}}\, ,\\
    &\bra{\zeta}\operatorname{K_+}\ket{\zeta'}
    =2\mu\,\displaystyle{\frac{(1-|\zeta|^2)^{\mu}(1-|\zeta'|^2)^{\mu}}{(1-\zeta^*\zeta')^{2\mu}}\,\frac{\zeta^*}{1-\zeta^*\zeta'}}\, .
\end{eqnarray}
\subsection{Coherent state path integral}

Equipped with these ingredients, we can now construct 
the imaginary-time path integral of the propagator $\bra{\zeta_{{\rm f}}}
e^{-i \operatorname{H}t_f/\hbar}\ket{\zeta_{{\rm i} }}$ between two coherent states,
where the Hamiltonian is given by
\begin{equation}
    \operatorname{H}=2\hbar\omega\operatorname{A} \, .
\end{equation}
After the usual procedure of constructing a path integral, see, e.g., Ref.\ \cite{KleinertBook}, we arrive at 
\begin{equation}
        \bra{\zeta_{f}}
       e^{-i \operatorname{H}t_f/\hbar} \ket{\zeta_{i}} = \int_{|\zeta|^2<1} \mathcal{D}(\zeta^*,\zeta)
        \exp{ \frac{i\mu}{\hbar}\int_0^{t_f}dt \biggl[\frac{\zeta^*(t)\dot{\zeta}(t) -\dot{\zeta}^*(t)\zeta(t)}{1-\zeta^*(t)\zeta(t)}\\
        +2i\omega \, \frac{1+\zeta^*(t)\zeta(t)}{1-\zeta^*(t)\zeta(t)}}  \, , 
        \label{path1}
\end{equation}
where the time-sliced path integral measure reads
\begin{equation}
   \mathcal{D}(\zeta^*,\zeta)=\prod_{n=1}^{N-1}\frac{d\zeta^*_{n} d\zeta_{n}}{\pi} \, \frac{2\mu-1}{(1-|\zeta_{n}|^2)^2}\,.
   \label{measure1}
\end{equation}
By employing the su(1,1)-Holstein-Primakoff transformation
\begin{equation}
    \kappa = \frac{\zeta}{\sqrt{1-|\zeta|^2}} \, ,
    \label{equ;Holstein-Primakoff}
\end{equation}
we recover from Eq.\ (\ref{path1}) the path integral of a harmonic oscillator with a rescaled frequency for the Grundberg-Hansson anyon pair in the form 
\begin{equation}
\hspace*{-0.2cm}    \bra{\kappa_{f}} e^{-i \operatorname{H}t_f/\hbar}
    \ket{\kappa_{i}} = \int \mathcal{D}(\kappa^*,\kappa)
    \exp{ \frac{i\mu}{\hbar}\int_0^{t_f}dt \biggl[\kappa^*(t)\dot{\kappa}(t) -\dot{\kappa}^*(t)\kappa(t)
    +2i\omega\kappa^*(t)\kappa(t)\biggr]}     \, .
\end{equation}
Moreover, the time-sliced path integral measure, Eq.\ (\ref{measure1}), assumes the form
\begin{equation}
    \mathcal{D}(\kappa^*,\kappa)=\prod_{n=1}^{N-1}d\kappa^*_{n} d\kappa_{n}\,\frac{2\mu-1}{\mu\pi} \, .
\end{equation}

Furthermore, we consider the case of a Grundberg-Hansson anyon pair with frequency $\omega$ coupled to a bosonic oscillator  with frequency $\omega_0$, which represents the basis of the considered system-anyon bath model. To this end, we consider a bilinear coupling between the Grundberg-Hansson anyon pair with the anyon operators $\operatorname{K}_{\pm}$ and the bosonic harmonic oscillator with operators $\operatorname{a}$, $\operatorname{a}^{\dagger}$. Thus, the resulting model Hamiltonian reads
\begin{equation}
    \operatorname{H} = \hbar\omega_0\operatorname{a}^{\dagger}\operatorname{a} + 2\hbar\omega \operatorname{A} + c\Bigl(\operatorname{a}^{\dagger}+\operatorname{a}\Bigr)\Bigl(\operatorname{K}_{-}+\operatorname{K}_{+}\Bigr) \, ,
\end{equation}
where $c$ denotes the strength of the bilinear coupling.
The propagator for system and bath variables then reads
\begin{equation}
     \bra{a_{\rm f}}\bra{\zeta_{{\rm f}}} e^{-i\operatorname{H}t_{\rm f}/\hbar} \ket{\zeta_{{\rm i}}}\ket{a_{\rm i}}= \int \mathcal{D}(a^*,a)\mathcal{D}(\zeta^*,\zeta)\exp{-\frac{i}{\hbar}\mathcal{S}\Bigl[\zeta^*(t),\zeta(t),a^*(t),a(t)\Bigr]} \, .
\end{equation}
Next, we split the action into two parts according to 
\begin{equation}
    \mathcal{S}\Bigl[\zeta^*(t),\zeta(t),a^*(t),a(t)\Bigr] = \mathcal{S}_{\rm S}\Bigl[a^*(t),a(t)\Bigr] + \Tilde{\mathcal{S}}\Bigl[\zeta^*(t),\zeta(t),a^*(t),a(t)\Bigr]\, ,
\end{equation}
where the action depending on the bosonic harmonic oscillator degrees of freedom reads
\begin{equation}
    \begin{split}
        \mathcal{S}_{\rm S} \Bigl[a^*(t),a(t)\Bigr]= \int_0^{t_f}d{t}\biggl[ \frac{i\hbar}{2}\Bigl[ a^*(t)\dot{a}(t)-\dot{a}^*(t)a(t)\Bigr]-\hbar\omega_0 a^*(t)a(t)\biggr]   \, . 
    \end{split}
\end{equation}
The other part then contains the information about the Grunberg-Hansson anyon pair and its coupling, i.e., 
\begin{eqnarray}
    \hspace*{-1.4cm}\Tilde{\mathcal{S}}\Bigl[\zeta^*(t),\zeta(t),a^*(t),a(t)\Bigr]&=& 2\mu\int_0^{t_f}dt \biggl\{\frac{i\hbar}{2}\,\frac{\zeta^*(t)\dot{\zeta}(t) -\dot{\zeta}^*(t)\zeta(t)}{1-\zeta^*(t)\zeta(t)}
        -\hbar\omega \, \frac{1+\zeta^*(t)\zeta(t)}{1-\zeta^*(t)\zeta(t)} \nonumber\\
        &&-c\Bigl[a(t)+a^*(t)\Bigr] \frac{\zeta^*(t)+\zeta(t)}{1-\zeta^*(t)\zeta(t)}\biggr\} \, .     
    \label{equ;Wirkungbad}
\end{eqnarray}
Using the su(1,1)-Holstein-Primakoff transformation, Eq.\ (\ref{equ;Holstein-Primakoff}), the action in Eq.\ (\ref{equ;Wirkungbad}) becomes
\begin{eqnarray}
   \Tilde{\mathcal{S}}\Bigl[\zeta^*(t),\zeta(t),a^*(t),a(t)\Bigr]&=& 2\mu \int_0^{t_f}dt \biggl\{\frac{i\hbar}{2}\,\Bigl[\kappa^*(t)\dot{\kappa}(t) -\dot{\kappa}^*(t)\kappa(t)\Bigr]
    +2i\hbar\omega\kappa^*(t)\kappa(t) \nonumber\\
      &  &-c\sqrt{1+\kappa^*(t)\kappa(t)}\Bigl[a(t)+a^*(t)\Bigr]\Bigl[\kappa^*(t)+\kappa(t)\Bigr]\biggr\}     \, .
    \label{equ:Wirkungbad}
\end{eqnarray}
Hence, the anyonic character leads to a non-polynomial coupling between the bosonic oscillator displacement and the anyon pair. This forms the basis of the system-anyon bath coupling term and can readily be generalized to the case of many anyon pairs forming a bath.
\end{appendix}

\end{document}